\documentstyle[sprocl,epsfig]{article}

\begin{document}
\title{FORMATION AND DECAY OF CHIRAL CONDENSATES \\
IN RELATIVISTIC HEAVY ION COLLISIONS}
\author{BERNDT M\"ULLER \\
	Department of Physics \\
	Duke University \\
	Durham, NC 27708-0305, USA}

\maketitle 

\abstracts{We discuss the various physics aspects of hypothetical, coherent excitations of the pion field caused by spontaneously generated local distortions of the chiral order parameter.  Such distortions, which may occur in the expansion of a hot fireball created in relativistic nuclear collisions, may be influenced or even triggered by the interaction with strong electromagnetic fields through the axial anomaly.  We suggest that semiperipheral collisions may provide the best environment for the formation of these disoriented chiral condensates.} 

\section{Introduction} 
Pion emission with an asymmetric charge ratio, $N(\pi^0) \neq \frac{1}{2} \left( N(\pi^+) + N(\pi^-)\right)$ may result from the decay of a coherent, isosinglet excitation of the pion field.\cite{Lam84,AR91} Such an excitation can be formally represented in the form \cite{HS71,EMS81} 
\begin{eqnarray}
|\varphi\rangle_{I=0} 
&=& N \int d\hat{n} \exp \left( \sum_k \varphi (k) \hat{n} \cdot
    {\vec a}^\dagger _k \right) | 0 \rangle 
 =\nonumber  \\
&=& \prod_k \sum_N \sqrt{ \frac{\varphi(k)}{{\rm sh}\varphi(k)} }
    \frac{\varphi(k)^N}{(2N+1)!} 
  \left( a^{(0)\dagger}_k a^{(0)\dagger}_k 
  - 2a^{(+)\dagger}_k a^{(-)\dagger}_k \right)
  | 0 )
\end{eqnarray}
Here ${\vec a}^\dagger_k $ denotes the isovector creation operator for a pion with momentum $k$.  Specific charge states are indicated by a superscript $(0,\pm)$ and $\varphi(k)$ is the coherent wavefunction in momentum space.  A simple geometrical argument in isospin space \cite{Bjorken} shows that the ratio of neutral to all pions contained in such a state follows the probability distribution 
\begin{equation} 
P(R) = \frac{1}{2\sqrt{R}} \quad \mbox{ where } \quad 
  R = N(\pi^0) /N(\pi). 
\end{equation} 
while the average charge ratio $\langle  R \rangle = \int^1_0 P(R)RdR$ has the expected value $\frac{1}{3}$, the distribution $P(R)$ differs strongly from the Gaussian 
\begin{equation} 
P_{\rm stat} (R) \sim \exp \left( - \frac{(R-1/3)^2}{2\sigma^2}  \right)
\end{equation} 
that holds for a statistical ensemble of pions with net isospin zero.  Essentially the same result is obtained for a coherent $I = 0$ state containing $2N$ pions, for which the probability of finding $2n$ neutral pions is 
\begin{equation} 
P^{(0)}_{2N, 2n}
= \left(\frac {N!}{n!} \right)^2 \frac{ (2n)!}{(2N+1)!} 2^{2(N-n)} \to 
\frac{ 1}{2\sqrt{Nn}}
\end{equation} 
in the limit $n, N \to \infty, R = n/N$ fixed.  One also finds strong correlations among the charge channels; in the same limit \cite {GGM93} 
\begin{equation} 
C_{00} \to \frac{4}{5}, \quad C_{0+} \to - \frac{2}{5}, \quad C_{++} \to \frac{1}{5}, \quad {\rm etc., } 
\end{equation}
where 
\begin{equation} 
C_{ab} = \frac{ \langle N(\pi_a) N(\pi_b)\rangle}
         {\langle N(\pi_a) \rangle \langle N(\pi_b) \rangle - 1} . 
\end{equation} 
Clearly, neutral and charged pions are anticorrelated, whereas pions of equal charge (especially neutral pions) are strongly correlated. 

\section{ Formation of DCC} 

What makes these formal results interesting in the observation \cite{Bjorken,RW93} that such coherent, isosinglet pion excitations could occur spontaneously in the de-excitation of a highly excited region of space, in which the chiral order parameter -- the quark condensate $\langle \overline{q}q\rangle$ -- has been restored to its symmetric value 
$\langle \overline{q}q\rangle = 0$.  As the order parameter returns to its equilibrium value $|\langle \overline{q}q \rangle | = f_\pi$, its orientation in isospin space may temporarily differ from the isoscalar, true equilibrium configuration
\begin{equation}
\frac{1}{\sqrt{2}} \langle \overline{u} u + \overline{d} d\rangle = f_\pi, 
\qquad  \langle \overline{u} d \rangle
= \langle \overline{d} u \rangle 
= \frac{1}{\sqrt{2}} \langle \overline{u} u 
- \overline{d}d \rangle 
= 0. 
\end{equation} 
Hence the name disoriented chiral condensate (DCC).  The formation of a locally homogeneous, albeit disorientated, region of chiral condensate is analogous to the growth of a domain of local magentization when a ferromagnet is quickly cooled (``quenched'') below the Curie temperature. 

This process can be conveniently analyzed in the framework of the linear sigma model.\cite{RW93}  Denoting the chiral order parameter field by $(\phi_0, {\vec \phi})$
\begin{equation} 
 \phi_0 = \langle \overline{q}q \rangle, \quad
{\vec\phi} = \langle \overline{q} {\vec\tau} q \rangle, 
\end{equation} 
one finds that a mode of momentum $k$ follows the equation 
\begin{equation} 
 \frac{\partial^2} {\partial t^2} {\vec \phi}_k (t) 
= \left( \lambda f^2_\pi - k^2 - \lambda \langle \phi(t)^2 \rangle \right) 
  {\vec \phi}_k(t).
\end{equation}
The order parameter grows exponentially when the factor in parentheses on the right-hand side is positive, i.e. 
$k^2 < \lambda (f_\pi^2 - \langle \phi^2 \rangle )$. 
This condition requires that 
\begin{enumerate} 
\item  the mode has a long wavelength and 
\item  that the expectation value and the fluctuations in the order parameter $\phi$ are small compared with $f_\pi$.  This condition is fulfilled after a quench. 
\end{enumerate}
Although the exponential growth affects modes with small $k$, it is not easy to grow spatially large domains, because the growth shuts itself off when $\langle \phi^2\rangle $ approaches the equilibrium value $f_\pi$.  One finds that domains are characteristically of size $r = (\sqrt{\lambda} f_\pi)^{-1} \approx 0.5$ fm for $\lambda$ = 20 demanded by sigma model phenomenology.\cite{GGP94}  It turns out that it is more favorable to consider a region of space in which some thermal fluctuations of the order parameter remain.  Denoting the effective potential as 
\begin{equation} 
    V_{\rm eff} (\phi)
 = - \frac{1}{2} \mu^2 \phi^2 + \frac{1}{4} \lambda \phi^4, 
\end{equation} 
where $\mu^2$ is a function of $T$, with 
$\mu^2(0) = \lambda f_\pi^2$ and $\mu^2 (T_c) = 0$, one finds that the domain growth is governed by the correlation function \cite{GM94} 
\begin{equation} 
C(r, t) 
= \langle \phi(0, t) 
          \phi(x, t) \rangle \to \exp \left(\mu t - \frac{\mu r^2}{2t} \right) 
\end{equation} 
where $r = |x|$.  As $T$ approaches $T_c, \mu$ becomes small, implying slower exponential growth, but also longer correlation lengths.  As growth stops after a time $t_f \sim \mu^{-1}$, the domain size is $\langle r^2 \rangle ^{1/2} \sim \mu^{-1}$. 

In the real world, since chiral symmetry is explicitly broken -- corresponding to the presence of an external magnetic field in the magnetic analog -- no disorientated domains will form when the system is annealed. 
Very small domains will form if the system is fully quenched.  
The ideal scenario must be somewhere in between.  It turns out that a three-dimensional expansion of a hot region of space filled with pion gas is close to the ideal conditions.\cite{Randrup}  As the region expands and cools below $T_c$, the long wavelength pionic modes enter the unstable region but the system stays close to the critical line $\mu = 0$.  This expectation is confirmed by numerical calculations.\cite{AHW95} 

\section{Decay of DCC}

The coherent pion excitation eventually decays into individual pions by decoherence of its individual components of different momentum due to residual interactions.\cite{BK92}  Consider a coherent state in the O(4) linear sigma model with explicit symmetry breaking 
\begin{equation} 
{\cal L} = \frac{1}{2} (\partial_\mu \phi)^2
- \frac{1}{4} \lambda(\phi^2- f_\pi^2)^2
- m^2_\pi f_\pi \phi_0, 
\end{equation} 
which has ``rolled down'' to a minimum of the potential along the chiral circle $|\phi| \approx f_\pi$.  Neglecting explicit symmetry breaking for a moment, the radial and angular directions at this field configuration define isoscalar $(\sigma)$ and isovector $(\pi)$ modes in the local environment of the DCC.  The decoherence of the coherent state is facilitated by the special form of the interaction between the radial and angular modes of oscillation.  Consider, for simplicity, the fate of a coherent radial oscillation in the true minimum $\phi = (f_\pi, {\vec 0})$ of the potential (12).  Writing $\phi_0 = f_\pi + \chi$ and ${\vec \phi} = {\vec \pi}$, the angular $(\pi)$ oscillations are governed by the equations \cite{MM95} 
\begin{equation} 
\left[ \frac{d}{dz^2} + A_\pi - 2 q_\pi \cos (2z) \right] {\vec \pi}_k(z) 
= 0
\end{equation} 
where the radial $(\sigma)$ oscillations are approximated by 
\begin{equation} 
{\chi}(t) = {\chi}_0 \cos (m_\sigma t + \alpha) 
\end{equation} 
with $m^2_\sigma = 2\lambda f_\pi^2$ and $A_\pi = 4(m^2_\pi + k^2)/m^2_\sigma, 
q_\pi = 2 {\chi}_0/f_\pi$.  The Lam\'e equation (13) has unstable, i.e. exponentially growing solutions in a wide range of parameters $(A_\pi, q_\pi)$. The unstable regions focus near $A_\pi = n^2 \quad (n = 1,2,\ldots )$ when $q_\pi \to 0$, but broaden as $q_\pi$ grows.

The quantum field theoretic description of this process has recently been worked out.\cite{HM97}  It turns out that the final state is described as a {\em squeezed} state of coherent pion pair excitations in the spontaneously broken chiral vacuum.  The classical equation (13) governs the amplitude functions determining the squeezed state.  The numerical results show a strong peak of the pion number in the momentum range where the Lam\'e equation (13) exhibits instability.  Strong charge and momentum correlations among pions are also found.  For pions of opposite momentum these correlations are dictated by those arising from the decay $\sigma \to \pi\pi$ while those among pions with nearly equal momenta are predominantly given by quantum correlations among identical particles (HBT correlations). 

\section{The Anomaly Effect} 

In the presence of strong electromagnetic fields, the Adler-Bell-Jackiw anomaly describes an additional interaction that affects only neutral pions.  The interaction has the form 
\begin{equation} 
\delta {\cal L} = - \frac{\alpha}{\pi f_\pi} ({\vec E}\cdot {\vec H}) \phi_3
\end{equation} 
shifting the minimum to an oblique position on the chiral circle given by 
\begin{equation} 
(\phi_3)_{\rm min} \equiv \frac{\alpha} {\pi f_\pi m^2_\pi} {\vec E} \cdot {\vec H}. 
\end{equation} 
This shift provides a bias in the dynamics of the pion field in the $\pi^0$ direction whose sign depends on the sign of the invariant ${\vec E} \cdot {\vec H}$.  For heavy ion collisions at a nonvanishing impact parameter $b \neq 0$, this sign differs between the regions of space above and below the scattering plane.\cite{MM96}

The effect of the anomaly is numerically small even in strong electromagnetic fields due to the presence of the factor $\alpha$ in (15).  However, its effect is not negligible because of the spatial homogeneity of the interaction (15) over distances $r \gg m^{-1}_\pi$.  Consider a highly energetic collision along the $z$-axis of two equal, heavy nuclei at an impact parameter $b = R_A$, where $R_A$ is the nuclear radius.  The pseudoscalar electromagnetic invariant is then given by 
\begin{equation} 
 {\vec E} \cdot {\vec H}
= \frac{2 Z^2 e^2 \gamma^2}{(4\pi)^2 R^3_1 R^3_2} \, {\vec x} \cdot ({\vec b} \times {\vec v}) 
\end{equation} 
where $R^2_{1/2} = \gamma^2 (z \pm vt)^2 + ({\vec x}_\perp \pm \frac{1}{2} {\vec b})^2$.  At high energy $(\gamma \gg 1)$ the value of ${\vec E} \cdot {\vec H}$ is strongly peaked near $t = 0$, and the anomaly interaction can be approximated by a quasi-instantaneous kick to the $\pi^0$ field at the moment of encounter of the two nuclei.  As noted above, the kick is small but spatially homogeneous over a region of the size of the nuclear radius $R_A$.  

\begin{figure}[htb]
\vfill
\centerline{
\begin{minipage}[t]{.47\linewidth}\centering
\mbox{\epsfig{file=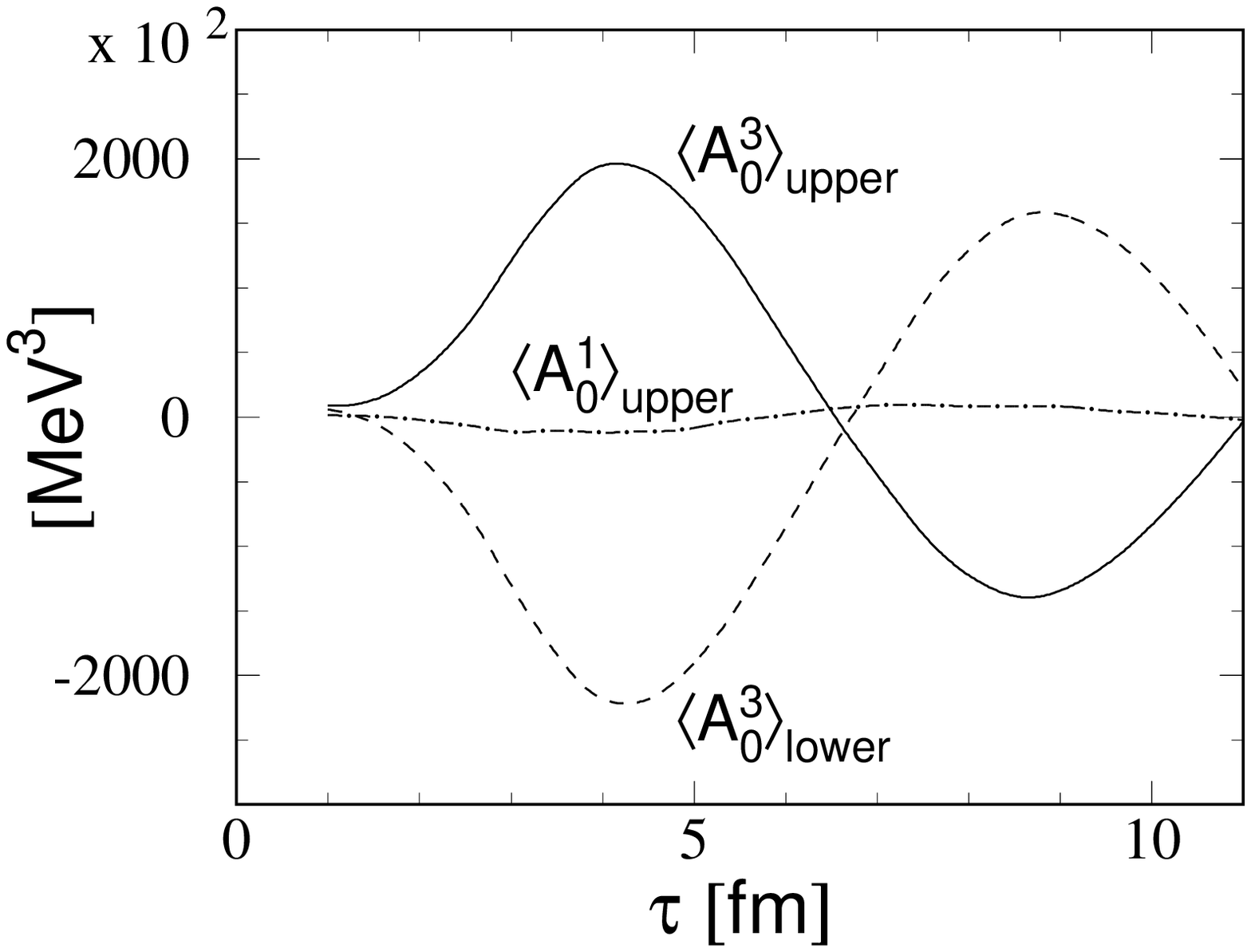,width=.99\linewidth}}
\caption{$\langle A_0^1\rangle_{\rm upper}$, $\langle A_0^3\rangle_{\rm upper}$, and $\langle A_0^3\rangle_{\rm lower}$ as a function of proper time. The calculation is for an anomaly strength $a_n=0.1$ and the average is taken over 10 events. [From ref.~\protect\cite{AMM98}]}
\label{fig-axial}
\end{minipage}
\hspace{.06\linewidth}
\begin{minipage}[t]{.47\linewidth}
\mbox{\epsfig{file=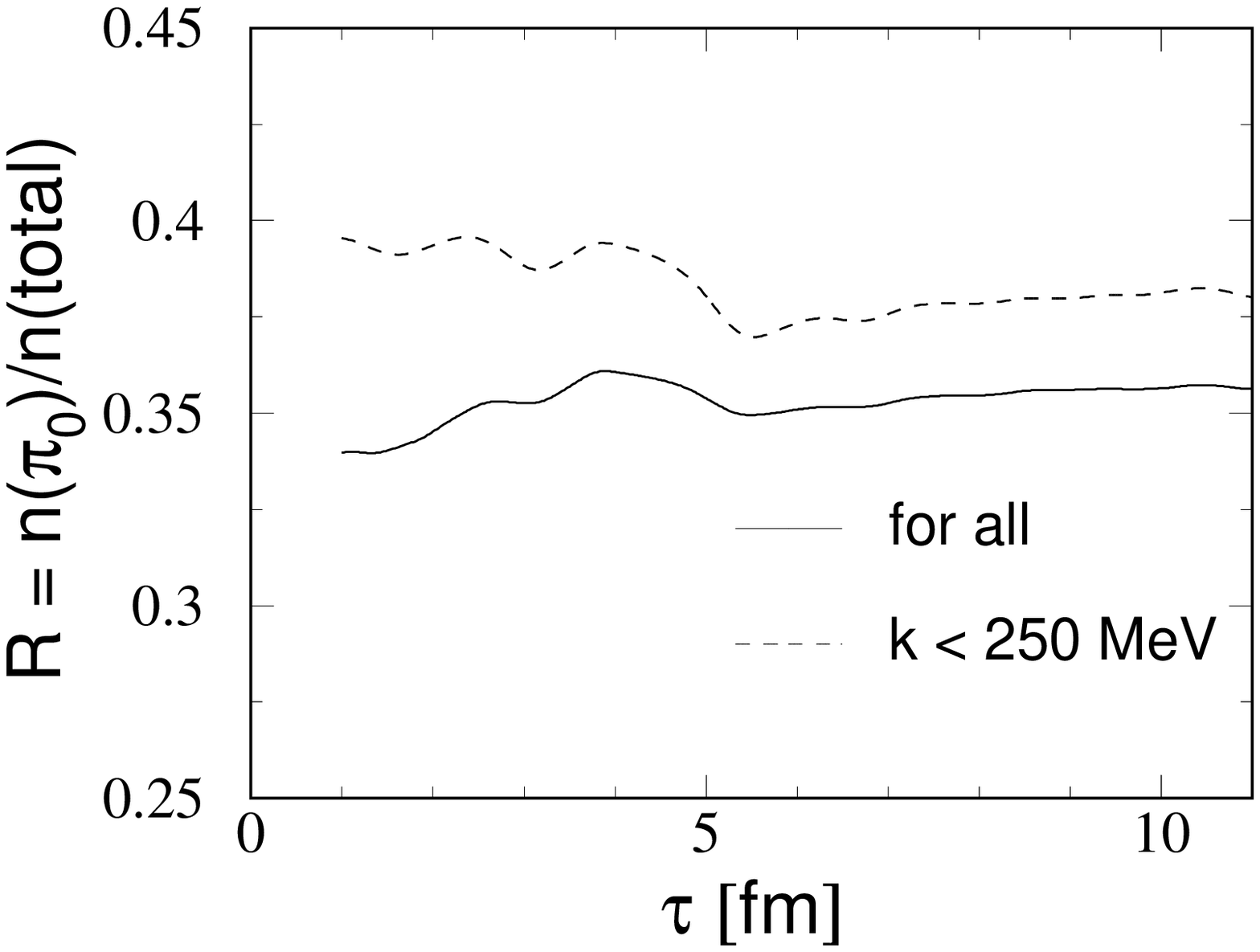,width=.99\linewidth}}
\caption{The solid line is the proper time evolution of $N(\pi^0)/N(\pi)$ averaged over 10 events. The dashed line is the same except that the momentum integration is limited to $k < 250$ MeV. [From ref.~\protect\cite{AMM98}]}
\label{fig-pion}
\end{minipage}}
\end{figure}

To investigate the effect of the anomaly kick on the formation of DCC domains, we have performed numerical simulations \cite{AMM98} of the linear sigma model in the modified ``quench'' scenario \cite{AHW95} with a schematic initial kick 
\begin{equation} 
\Delta {\dot \phi}_3 = {\rm sgn} (y) a_n m^2_\pi, 
\end{equation} 
where $y = 0$ denotes the scattering plane.  According to (17) the value of $a_n$ should lie in the range 0.05 -- 0.1, for a value of $\gamma = 100$ corresponding to nuclear collisions at RHIC.  Our simulations show that the anomaly kick has a surprisingly large effect on the time evolution of the $\pi^0$ field, although the initial kick is tiny when compared with the fluctuations of the pion field on length scales of order $m^{-1}_\pi$.  The effect and the asymmetry with respect to the scattering plane is best revealed by looking at the time-like component of the neutral axial current, $A^3_0$, as a function of time (see Fig.~\ref{fig-axial}).  While the sign of $A^3_0$ varies over time, the sign always differs between the upper and lower half-space.  We also found that no other component of the axial or the vector current shows a similar behavior, and we have confirmed that the effect remains unaffected by a restriction of the anomaly kick to a finite region of space. 

It appears difficult to detect this polarization of the axial charge directly, because pionic observables are usually not sensitive to the sign of $A_0^3$.  However, we have found that the effect is sufficiently strong to slightly distort the isospin symmetry of DCC formation.  This effect could be large enough so that it may be observable in a large sample of events, by comparing the average number of produced neutral pions to the average number of charged pions (see Fig.~\ref{fig-pion}).  The enhancement is concentrated at small momenta $(k < m_\pi)$ and hence may be extracted by averaging the $\pi^0/\pi^\pm$ ratio over many events for high and low pion momenta, respectively. 

\section{Summary} 

Coherent excitations of the chiral order parameter field, i.e.\ the physical pion field, in high energy heavy ion collisions can generate unusual distributions of the charge ratios of produced pions as well as peculiar pion charge correlations.  The three-dimensional expansive cooling of an equilibrated fireball with initial temperature $T > T_c$, the critical temperature of the chiral phase transition, provides a ``natural'' and plausible mechanism of such excitations in the form of domains of disoriented chiral condensate (DCC).  Since the condition of approximately vanishing isospin is crucial for the phenomena discussed here, they are most likely to occur in the nearly baryon-free central rapidity region in nuclear reactions at RHIC. 

The electromagnetic axial anomaly enhances the spectral power of the $\pi^0$-excitations at small momentum and manifestly breaks the $\pi^0/\pi^\pm$ symmetry.  This effect may result in a neutral pion ratio $N(\pi^0)/N(\pi) > \frac{1}{3}$ in this region of phase space.  The anomaly effect could also lead to an overall enhancement of the formation of extended DCC domains in semiperipheral collisions of very heavy nuclei at RHIC or LHC. 

\section*{Acknowledgement} 

I am indebted to M. Asakawa and H. Minakata for their collaboration and many insightful discussions.  This work was supported in part by the U.S.\ Department of Energy under Grant No. DE-FG02-96ER40945.

\end{document}